\documentclass[12pt]{article}

\usepackage[paper=a4paper, margin=2cm]{geometry}
\usepackage[russian,english]{babel}
\usepackage[cp1251]{inputenc}
\usepackage{amsmath}

\usepackage[logos]{amsbib}

\begin{document}

\date{}
\renewcommand{\refname}{References}

\author{A.~Yu.~Samarin
}
\title{Quantum jump mechanics}

\maketitle

\centerline{Samara Technical State University, 443100 Samara, Russia}
\centerline{}

\abstract{The change with time of the system consisting of the quantum object and the macroscopic measuring instrument is described on the base of the uniform dynamic law, which is suitable both evolution and reduction processes description. It is the integral wave equation with kernel in the form of a path integral. It is shown, that wave function collapse is the specific transformation which is fundamentally differ from Shr\"odinger's evolution. Specifically, a formal cause of the collapse is a local time derivative (infinite large) of the potential energy. Such transformation can not be described using mathematical apparatus of conventional quantum mechanics.



\begin {center} \textbf{1. Introdution}\\
\end {center}
  The wave equation describing evolution of quantum objects \footnote{Hereinafter we consider the nonrelativistic case only.} does not imply in itself the notion of "observable". This notion formally arises, when we consider the part of this equation as Hermitian Hamilton's operator \footnote{It is the result of joint analysi of the wave equation and the eigenvalue problem of Hermitian Hamilton's operator}. This approach reduces the set of all possible solutions of the wave equation to the set of the stationary wave function only\footnote{Hereinafter we consider  coordinates wave functions.}. This assumption --- another postulate of conventional quantum mechanics --- eliminates out of consideration those states of the quantum objects satisfying the wave equation, that are not eigenfunctions of the energy operator\footnote{This postulate has to do with any observable. It is uniformly expressed in the postulate of the Hilbert space~\cite{bib:Aa}.}. That is, there exists only those states that can directly cause macroscopic changes in the measurer. On the other hand, the wave function is a unique direct mathematical image of the quantum object. Any observable is not only attribute of a quantum object, it is the result of the measuring process. Taking into account this peculiarity of the observable, we may doubt that corresponding wave functions describe all possible states of a quantum object. The existence of virtual quantum processes, that are cannot be considered in detail due to the uncertainty of energy, indirectly confirms this doubt. In order to understand, what kind of the wave function is authentic, it is necessary to study the dynamics of the measurement process. But a problem exists: the measuring process results in "wave function reduction" or "quantum jump"~\cite{bib:Aa,bib:Ab}. In contrast to the evolution process this formal mathematical procedure has no interpretation as a causal chain of events. Moreover exact understanding does not exist, what is it in effect? This problem generated the nonphysical idea to include into consideration of quantum processes the mind of the observer~\cite{bib:Aa}. Even if to ignore various problems caused by this assumption, it can not be taken if we want to operate within a physical theory (concept of mind is not a physical concept). The approach presented in~\cite{bib:Ac} has generated numerous attempts to avoid this problem in principle~\cite{bib:Ad,bib:Ae,bib:Af}. The many universes interpretation of quantum mechanics postulates the simultaneous existence of alternative realities, which exist independently of each other. Any specific result of the collapse, in this theory, determines physical pattern of the corresponding reality. In this case the collapse is not a physical process it is a manner to assay the actual reality, that we live in. But these universes are not connected with each other physically and, therefore, all other universes can not effect on our universe, i.e. they are not physical realities for our world by definition. Thus, all the universes (besides one) are not more then a fantasies. This interpretation becomes explicitly absurd if to suppose validity of the universal causal principle. Consider the state of the classical motion of the tossed coin in form of a mixed state. In this case, the fundamental possibility to describe the system dynamically, does not allow to interpret the mixed state as a reality consisting of two universes which are correspond to the alternative results. This mixed state is not more than a demonstration of the fact, that our information about the physical system is not complete.

 Most of the difficulties of interpretation of quantum mechanics are generated by the lack of causality principle. In order to overcome this difficulties, it is necessary to postulate the deterministic equation describing both Schr\"odinger's evolution and the collapse. Since the collapse is the nonlocal~\cite{bib:Ag, bib:Ai} and nonlinear~\cite{bib:Aj} phenomenon, then Schr\"odinger's equation, which is local and linear, can not be considered as the base of such equation. It have to be replaced by other dynamics law having nonlocal and nonlinear properties too. In addition to the mechanism of linear and local evolution of the wave function this law have to contain the nonlinear and nonlocal transformation mechanism of the wave function. Such properties has the integral wave equation with the kernel in the form of a path integral~\cite{bib:Ak,bib:Al}\footnote{Evolution of the wave function described by the mechanical motion corresponding to the virtual paths is the local linear process. It is Schr\"odinger's evolution~\cite{bib:Al}. But if a physical process produces changes with the time the measure of the set of the virtual paths passing through a local spatial domain (for example, producing temporal changes of the potential energy values in this spatial domain), then this measure variation causes the simultaneous wave function transformation in whole space due to the normalization conservation, i.e. this transformation is nonlocal. In the case when the process dynamics depends on the wave function, this transformation is nonlinear too.}. Search of the physical situation, in which such process is initiated in the form of the spatial collapse, is the goal of this paper.

 \begin {center} \textbf{2. The main principles of the approach}\\
\end {center}

  Consider a closed system composed of the quantum particle\footnote{Hereinafter, for simplicity, all quantum objects will be considered in the form of quantum particles. The quantum particle is an indivisible quantum object, that is transformed into a mass point as a result of the collapse.} and the measuring instrument. The wave function of this system has to transform in accordance with the universal law of dynamics including both quantum evolution and reduction of the wave function~\cite{bib:Am,bib:An}. This means that a measurer has to be described as a usual physical object obeying the laws of quantum mechanics. Therefore, at first, it is necessary to define those physical properties of a measurer, that differ it from all others physical objects. Ultima analysi readings of any measuring instrument are macroscopic. On the other hand, an interaction of the quantum object with the measuring instrument is described by quantities having the order of magnitudes, which is typical for quantum processes. Thus, the measuring process has to contain amplification some of these quantities up to the macroscopic level\footnote{In the case of the interaction of a photon with an eye, the amplification process proceeds in the eye.}. \emph{It is the amplification process is considered in this paper as the cause of the wave function collapse.}

 We mentioned earlier, that Schr\"odinger's evolution is a local process, but the collapse is nonlocal. Therefore the universal law of dynamics has to possess both these properties. Integral wave equation with the kernel in the form of a path integral can be considered as such law. Indeed, the evolution of the wave function, can be represented as a continuous medium motion~\cite{bib:F} along the virtual paths \footnote{Indeed, if the collapse is a real nonlocal physical process, then the wave function can be considered as the distribution of the density of a stuff of which the world is made (if the collapse exists really, then the spreading wave packet is not a difficulty for such interpretation of the wave function).}. The motion of individual points of this continuum is completely determined by the values of the physical quantities at the space locus, where this individual point locates at a given time. I.e. in this case, we have a local dynamics. On the other hand, if the composition of the quantum system is fixed, then normalization of the wave function conserves in any processes. This normalization as a mathematical procedure reflects the possibility of nonlocal processes of the wave function transformation. Indeed, a time change of the potential energy in the local region of space results in corresponding change of the measure of the virtual paths passing through this region~\cite{bib:An}. This last change immediately causes the wave function transformation in all space. That is, a local time derivative of the potential energy generates the nonlocal wave function transformation.

  In addition, the integral wave equation allows to consider a macroscopic body in terms of quantum evolution~\cite{bib:Fa}. All these resources of the integral wave equation make it a unique instrument for description of the wave function collapse.

  \begin {center} \textbf{3. Mathematical representation of the measuring process}\\
\end {center}

Consider the wave function collapse in the form of quantum particle localization in physical space. Evolution of the wave function is considered as a one-dimensional mechanical motion. Measurer encloses $N $ actives elements (hereinafter grains) interacting with the quantum particle-object. These grains possess the amplification property: interaction of the quantum particles of the grain with the particle-object can generate in this grain a macroscopic process (registering process)\footnote{The amplification property, mentioned above, is an attribute of the grains and construction of other parts of the measurer. Specific form of the construction has no matter for collapse considering.}.

Let the grain size be so small, that its position in space can be determined by the unique coordinate $Y^{j} $ (superscript indicates the number of the grain)\footnote{As localization of the quantum particle in space go with coordinate measuring, then this condition has to be hold if we want to provide accuracy of the coordinate measuring.}. For validity of this assumption, it is necessary, that wave function of the particle-object has the same values for volume of the grain.

We call the active particles those the grain particles, that interact directly with the particle-object. Denote by $z^{j,k}_{t} $ the spacial coordinates of the active particles (the first superscript is a number of the grain; the second is a number of the active particle of this grain; the subscript indicates the instant of time corresponding to the value of this coordinate). Let $M $ be the total number of active particles in the grain. In ~\cite{bib:Fa} is shown that classical mechanical motion of the center of mass characterizes system of quantum particles as a macroscopic body. The main idea of this paper is that the action functionals on the center of mass virtual paths increases up to a macroscopic value in the measuring process and this increase generates the wave function collapse of the particle-object. Taking into account this idea, consider the wave function of the system in the form of a function of the particle-object coordinates $x $, coordinates of the centers of mass of the active particles collections of the grains $X^{j}=\frac{\sum\limits_{j=1}^{n}(m^{k}z^{j,k})}{\sum\limits_{j=1}^{n}(m^{k})} $ and coordinates characterizing active particles relative motion  $\xi^{j,k}=z^{j,k}-X^{j} $. Time dependence of such wave function is determined by the integral wave equation for the configuration space~\cite{bib:An}:
\begin{multline}
\label{eq:math:ex1}
    \Psi_{t_{2}}(x_{2},X_{2}^{1}, \dots ,X_{2}^{N},\xi_{2}^{1,1}, \dots ,\xi_{2}^{N,M-1})\\
    =\idotsint K_{t_{2},t_{1}}(x_{2},X_{2}^{1}, \dots ,X_{2}^{N},\xi_{2}^{1,1}, \dots ,\xi_{2}^{N,M-1},x_{1},X_{1}^{1}, \dots ,X_{1}^{N},\xi_{1}^{1}, \dots ,\xi_{1}^{N,M-1})\\
    \times\Psi_{t_{1}}(x_{1},X_{1}^{1}, \dots ,X_{1}^{1},\xi_{1}^{1,1}, \dots ,\xi_{1}^{N,M-1})\,dx_{1}\,dX^{1}_1\cdots\,dX^{N}_1 d\xi^{1,1}_1\cdots\,d\xi^{N,M-1}_1,
\end{multline}

where
\begin{multline}
\label{eq:math:ex2}
    K_{t_{1},t_{2}}(x_{2},X_{2}^{1}, \dots ,X_{2}^{N},\xi_{2}^{1,1}, \dots ,\xi_{2}^{N,M-1},x_{1},X_{1}^{1}, \dots ,X_{1}^{N},\xi_{1}^{1}, \dots ,\xi_{1}^{N,M-1})\\
    =\int\exp{\frac{i}{\hbar}s_{1,2}[x(t)]}\idotsint\prod\limits_{j=1}^{N}\Biggl(\exp{\frac{i}{\hbar}S_{1,2}^{j}[X^{j}(t)]}\\
    \times\idotsint\prod\limits_{k=1}^{M-1}\biggl( \exp{\frac{i}{\hbar}S_{1,2}^{jk}[\xi^{j,k}(t)]}\,[d\xi^{j,k}(t)]\biggl)\,[dX^{j}(t)]\Biggl)\,[dx(t)]
\end{multline}
is a transition amplitude in the form of a continual integral~\cite{bib:ASA}.
In general, interaction energy of the particle-object with active particle of the grains in last expression have to enter in action functional of the relative motion $S_{1,2}^{jk}[\xi^{j,k}(t)] $\footnote{The interaction energy is attribute of the paths of the particle-object and active particles of the grains in the configuration space. If to express, formally, corresponding action functional as a sum of the action functionals of the particle object $s_{1,2}[x(t)] $, the active particles center of mass $S_{1,2}^{j}[X^{j}(t)] $ and relative movement $S_{1,2}^{jk}[\xi^{j,k}(t)] $, then this energy have to be taken into account only in one of them.}. But if wave function of the particle-object has the same values in the volume of the grain, then this energy does not depend on the variables $\xi^{j,k}$ and can be taken into account in action functional of the active particles center of mass $S_{1,2}^{j}[X^{j}(t)] $. This means that transition amplitude \eqref{eq:math:ex2} can be expressed in the form of
\begin{multline*}
    K_{t_{1},t_{2}}(x_{2},X_{2}^{1}, \dots ,X_{2}^{N},\xi_{2}^{1,1}, \dots ,\xi_{2}^{N,M-1},x_{1},X_{1}^{1}, \dots ,X_{1}^{N},\xi_{1}^{1}, \dots ,\xi_{1}^{N,M-1})\\= P_{t_{1},t_{2}}(x_{2},X_{2}^{1}, \dots ,X_{2}^{N},x_{1},X_{1}^{1}, \dots ,X_{1}^{N})Q_{t_{1},t_{2}}(\xi_{2}^{1,1}, \dots ,\xi_{2}^{N,M-1},\xi_{1}^{1}, \dots ,\xi_{1}^{N,M-1}),
\end{multline*}
i.e. the interaction process does not depend on the relative movement of the active particles. Then the interaction process can be described by the wave function
\begin{multline}\label{eq:math:ex3}
    \Psi_{t_{2}}(x_{2},X_{2}^{1}, \dots ,X_{2}^{N})\\
    =\idotsint K_{t_{2},t_{1}}(x_{2},X_{2}^{1}, \dots ,X_{2}^{N},x_{1},X_{1}^{1}, \dots ,X_{1}^{N})\\
    \times\Psi_{t_{1}}(x_{1},X^{1}_{1} \dots X^{N}_{1})\,dx_{1}dX^{1}_1\cdots\,dX^{N}_1,
\end{multline}
where
  \begin{multline}
K_{t_{2},t_{1}}(x_{2},X_{2}^{1}, \dots ,X_{2}^{N},x_{1},X_{1}^{1}, \dots ,X_{1}^{N})\\
 =\int\exp{\frac{i}{\hbar}s_{1,2}[\gamma]}\Biggl(\idotsint\prod\limits_{j=1}^{N}\exp{\frac{i}{\hbar}S_{1,2}[\Gamma^{j},\gamma]} \,[d\Gamma^{j}]\Biggl)\,[d\gamma]
 \label{eq:math:ex4}.
\end{multline}
 In this expression, we use the notations  $\gamma\equiv x(t) $ and  $\Gamma^{j}\equiv X^{j}(t) $.
 Potential energy of the interaction of the particle-object with the grains depends on the variables $x $ and $X $. It links the particle-object and ensembles of the active particles into a united system, described by the entangled wave function.

 The process in measurer initiating by the interaction with the particle-object have to be amplified up to a macroscopic level. Now consider haw this amplification effects on the forms of expressions  \eqref{eq:math:ex3} and \eqref{eq:math:ex4}. Transition amplitude \eqref{eq:math:ex4} is determined by space of the virtual paths. This space is formed by the paths passing through the spacial domain where the quantum object can be detected ~\cite{bib:ASA, bib:ASS}\footnote{In the context of this paper the term "be detected" has to be replaced by the term "be"}. Rigorous mathematical criterion of this choice of the paths set is not currently available. But it can be introduced using results of~\cite{bib:F}. If the real path of an individual particle of the quantum continuous medium are determined by the minimum action principle, then, taking into account time homogeneity, energy of this particle has to be conserved. I.e. the individual particles can only locate in the region of space allowed by energy conservation law. This statement does not contradict the existence of the tunnel-effect. Really, energy of the individual particles collection has uncertainty, whereas the energy of the single particle (unobservable quantity) has an exact value.
 Thus, if energy of the active particles changes, then the space of the virtual paths can vary too due to the changing of the region of space allowed for mechanical motion of the individual points.

 In order to assay the role of this space transformation in the measurement process, consider following simple interaction model. Without interaction with the particle-object the active particles of the grains locates in identical potential wells. Coordinate of their center of mass is fluctuates in infinitesimal neighborhood of the equilibrium position \footnote{In this case the active particles are in finite stochastic motion. Number of these particles in any grain has a macroscopic value.}. Formally, this means that the path integral describing the relative movement in \eqref{eq:math:ex2} tent to zero, when active particle number tend to infinity. This means that the center of mass is quiescent.

  General form of the particle-object interaction with the active particles can be ascertained using the integral wave equation. Detail analysis of this equation producing in~\cite{bib:F} disclosed, that it results in the variational principle:
  \begin{equation*}
  \langle\delta S[\Gamma]\rangle=0,
  \end{equation*}
  where
  \begin{equation*}
  \langle\delta S[\Gamma]\rangle=\int S[\Gamma]\exp\frac{i}{\hbar}S[\Gamma][d\Gamma],
  \end{equation*}
  $\Gamma $ is a virtual path in the configuration space, $S[\Gamma] $ is a classic action functional for the virtual path. Following classical mechanics we obtain
  \begin{equation*}
  \biggl\langle\frac{d}{dt}\frac{\partial L}{\partial\dot{X}}+\frac{\partial L}{\partial X} \biggl\rangle=0,
  \end{equation*}
  where $X $, $\dot{X}$ are generalized coordinate and generalized velocity of the individual particles of the center of mass positions field. By $x $ and $\dot{x} $ denote generalized coordinate and generalized velocity of the individual particles of the particle object matter field. Then the interaction between individual particles of the particle-object continuum and individual particles of the continuum of the center of mass positions is described by the equation
  \begin{equation*}
  m\langle\ddot{X}\rangle+\frac{\partial \langle U(x,X)\rangle}{\partial X}=0,
  \end{equation*}
  where
  \begin{equation*}
  \langle\ddot{X}\rangle=\int \ddot{X}[\Gamma^{j}]\exp\frac{i}{\hbar}S[\Gamma][d\Gamma];
  \end{equation*}
   \begin{equation*}
  \langle U(X,x)\rangle=\int U\biggl(X(t)-x\biggl)\exp\frac{i}{\hbar}S[\Gamma][d\Gamma].
  \end{equation*}
   The last term of this equation is the cause of evolution of the individual point motion (the first term). I.e. the last term is the quantum analog of the classical notion of mechanical force. The last term has to be integrated over all individual particles of the particle-object continuum (the first term does not depend on object coordinates). As all individual particles are identical, in accordance with considered here interpretation, the wave function of the particle-object is a weighting function. Thus, for the last term, we have
   \begin{equation*}
  \langle U_{t}(X-x)\rangle=\int\Biggl(\int U\biggl(X(t)-x\biggl)\exp\frac{i}{\hbar}S[\Gamma][d\Gamma]\Biggl)\Psi_{t}(x)dx.
  \end{equation*}
  In accordance with conventional quantum mechanics any wave function is a superposition of the stationary state wave functions. Considered here interpretation does not state this. In order to escape the conflict with the conventional quantum mechanics it is necessary to suppose that the measuring processes are possible, when particle-object is only in stationary states. Let the particle-object is in a stationary state. Taking into account that the size of any grain is small and the wave function has the same value in its volume, we obtain
  \begin{equation}
  \langle U_{t}(Y^{j}-x)\rangle=\exp \biggl(-\frac{i}{\hbar}Et\biggl)\int\Biggl(\int U\biggl(Y^{j}(t)-x\biggl)\exp\frac{i}{\hbar}S[\Gamma][d\Gamma]\Biggl)\psi(x) dx.
  \label{eq:math:ex5}
  \end{equation}
This expression implies an ideal measurement, when wave function of an object does not depend on the measurer state.

Thus, the potential energy \eqref{eq:math:ex5} has a periodic time dependence. Therefore its gradient (mechanical force) has a periodic time dependence too.
Then, when the measuring process has begun, all active particles of a grain move around periodically with the same phase of the external force. Formally, according to the considered model, this means that the function $X(t)$ is a harmonic ratio. Energy of this oscillation increases with the time. The rate of this increase is in proportion to square of modulus of the external force \footnote{Relations between the physical quantities characterizing quantum objects (matrix elements~\cite{bib:Al}) has a classical form~\cite{bib:F} and we have the situation identical to classical forced oscillation.} and, taking into account \eqref{eq:math:ex5}, to the quantity $|\psi (x)|^{2}$ for $x=Y^{j}$.

Thus, we have in-phase oscillating movement of the active particles of every grain. Therefore, the change with time of the center of mass coordinate $X^{j}(t)$ has the form of the oscillating movement with increasing energy too. Let all individual particles of all active particles of a grain have the same energy before the interaction with the particle-object. Then the oscillating movement of their center of mass can be considered as a movement in the potential well having a specified depth. As soon as value of the center of mass energy exceeds the value of this depth, the range of possible values $X^{j}$ changes\footnote{I.e. the space of the virtual paths  $X^{j}(t)$ varies.}. Formally this means that in the action functional $S_{1,2}\Gamma^{j} $  in \eqref{eq:math:ex4}, new term $\int U^{j}_{A}(X^{j}dt)$ arises ($U_{A}$ is the potential energy of the active particles in external field). Denote by $t^{j}$ corresponding instant of time for the grain $j$. Then for $t>t^{j}$ we have
\begin{equation}
    S_{12}[\Gamma^{j},\gamma]=\int\limits_{t_{1}}^{t_{2}}\biggl(T^{\Gamma^{j}(t)}(\dot{X^{j}})-U_{q}^{j}(x,X^{j})-U_{A}^{j}(X^{j})h(t-t_{k})\biggl) \,dt.
\label{eq:math:ex6}
\end{equation}
In this expression $T^{\Gamma^{j}(t)}(\dot{X^{j}}) $ denotes the kinetic energy of the center of mass of the active particles of the grain  $j $; $h(t-t_{k}) $ is the Heaviside function.
As the grains are macroscopically identical, then the form of the energy $U^{j}_{A}(X^{j})$ does not depend on the number of the grain $j$.  Thus we have    $U^{j}_{A}(X^{j})=U_{A}$. Denote by $\varepsilon$ the time interval,  with the elapsing of which action functionals on the paths of all microscopic processes becomes negligible small compared with the quantity $U_{A}\varepsilon$.  The quantity $\varepsilon$ has the order of magnitude $\hbar/U_{A}$. It is practically infinitesimal value. The time $t^{j}$ entering in \eqref{eq:math:ex6} has a statistical straggling for the grains. Let minimum of these values corresponds to the grain $k$. Consider the transformation of the system wave function during the time interval $\varepsilon=t^{k}_{4}-t^{k}_{3}$. Before the time $t^{k}_{3}$ the energy $T^{\Gamma^{k}}$ has a microscopic order of magnitude, i.e. $T^{\Gamma^{k}}\ll U^{A}$. This inequality relationship conserves for the interval $\varepsilon$ \footnote{Really, kinetic energy of the macroscopic process equals zero at the time $t^{k}_{3}$. In correspondence with Newton's law, we have $T_{A}\sim\frac{1}{2M}\biggl(\frac{dU_{A}}{dX}\biggr)^{2}\varepsilon^{2}$. The mass $M$ and the force $\frac{dU_{A}}{dX}$ have macroscopic orders of magnitudes, therefore $T_{A}$ has the second infinitesimal order.}. Thus for the action functional $S^{k}(\Gamma)$ at the time $t^{k}_{4}$, we have the value $-U_{A}\varepsilon$.

 Since the quantity $\varepsilon$ is infinitesimal, we can neglect the spatial motion of all considered objects in this interval. Then the collection of all virtual paths of the system, corresponding to the grain $k$ at the time $t^{k}_{3}$, is the same at time $t^{k}_{4}$. Denote by $\Omega^{I}$ region of the configuration space corresponding to the grain $k$. Denote by $\Omega^{II}$ the rest of the configuration space. Then for corresponding transition amplitudes, we obtain
 \begin{equation}
K_{t_{4},t_{3}}(x_{4},X_{4}^{k},x_{3},X_{3}^{k})=\exp\biggl(-{\frac{i}{\hbar}U_{a}\varepsilon}\biggl)\Biggl(\int\exp{\frac{i}{\hbar}s_{34}[\gamma^{I}]}\,[d\gamma^{I}]\Biggl)\delta(X_{4}^{k}-X_{3}^{k}),
\label {eq:math:ex7}
\end{equation}
for $\Omega^{I} $ and
\begin{multline}
 K_{t_{4},t_{3}}(x_{4},X_{4}^{1}, \dots ,X_{4}^{k-1},X_{4}^{k+1}, \dots ,X_{4}^{N},x_{3},X_{3}^{1}, \dots ,X_{3}^{k-1},X_{3}^{k+1}, \dots ,X_{3}^{N})=\\
 =\int \exp{\frac{i}{\hbar}s_{34}[\gamma^{II}]}\prod\limits_{j=1,j\neq k}^{N}\Biggl(\int\exp{\frac{i}{\hbar}S_{3,4}^{j}[\Gamma^{j}]} \,[d\Gamma^{j}]\Biggl)\,[d\gamma^{II}],
\label{eq:math:ex8}
\end{multline}
for $\Omega^{II} $. Superscript $I$ of the particle-object path denotes the paths passing through the volume of the grain $k$ at the time $t_{3}$. Taking into account consistency of the virtual paths collections in \eqref{eq:math:ex7} and \eqref{eq:math:ex8}, for wave function normalization we have:
\begin{multline}
\idotsint\limits_\Omega\biggl|\Psi_{t}(x,X^{1}, \dots ,X^{n})\biggl|^{2}\,dx\,dX^1\cdots\,dX^N=\\
=\idotsint\limits_{\Omega^{I}}\biggl|\Psi_{t}(x,X^{1}, \dots ,X^{n})\biggl|^{2}\,dx\,dX^1\cdots\,dX^N+\\
+\idotsint\limits_{\Omega^{II}}\biggl|\Psi_{t}(x,X^{1}, \dots ,X^{n})\biggl|^{2}\,dx\,dX^1\cdots\,dX^N=1
\label{eq:math:ex9}.
\end{multline}

The changes with time of amplitudes \eqref{eq:math:ex7} and \eqref{eq:math:ex8}, which determine terms in \eqref{eq:math:ex9}, are independent\footnote{These amplitudes determine only relative values of the terms. Absolute magnitude of the terms in \eqref{eq:math:ex9} have to satisfy normalization requirement}. In order to ascertain the result of this change it is useful to transform the expressions for this amplitudes into the real form.

\begin {center} \textbf{3. The wave function normalization in the real form}\\
\end {center}
Let us introduce the complex time $t=\tau\exp i\varphi$ and transform the path integrals in \eqref{eq:math:ex7} and \eqref{eq:math:ex8} into a real form ~\cite{bib:ASS}. Such technique is used in order to specify measure of the path integral. In the last case the real time transforms into the complex form in accordance with expression $t=\tau\exp\bigl(-\frac{\pi}{2}\bigr)$ i.e. $t=-i\tau $ and chain of events reverses. It is acceptable for reversible mechanical motion, which generates the path integral measure. But it is not acceptable under consideration of the irreversible collapse. Thus, we have to use the time transformation according to $t=\tau\exp\frac{\pi}{2}$. \footnote{Under such transformation the continual integral measure can be specified using analytic continuation of the Wiener measure trough the axis of the real time.}. As the result of this transformation for the transition amplitudes we have
\begin{multline}
K_{\tau_{4},\tau_{3}}(x_{4},X_{4}^{k},x_{3},X_{3}^{k})=\\
 =\exp\biggl({\frac{1}{\hbar}U_{a}\varepsilon}\biggl)\Biggl(\int\exp{\biggl(-\frac{1}{\hbar}s_{34}[\gamma^{I}]\biggl)}\,[d\gamma^{I}]\Biggl)\delta(X_{4}^{k}-X_{3}^{k});
\label{eq:math:ex10}
\end{multline}

\begin{multline}
K_{\tau_{4},\tau_{3}}(x_{4},X_{4}^{1}, \dots ,X_{4}^{k-1},X_{4}^{k+1}, \dots ,X_{4}^{N},x_{3},X_{3}^{1}, \dots ,X_{3}^{k-1},X_{3}^{k+1}, \dots ,X_{3}^{N})=\\
 =\int\exp\biggl(-\frac{1}{\hbar}s_{34}[\gamma^{II}]\biggl) \Biggl(\idotsint\prod\limits_{j=1,j\neq k}^{N}\exp{\biggl(-\frac{1}{\hbar}S_{3,4}^{j}[\Gamma^{j}]}\biggl) \,[d\Gamma^{j}]\Biggl)\,[d\gamma^{II}],
\label{eq:math:ex11}
\end{multline}
In the last expressions the quantities $s[\gamma], S[\Gamma^{j}] $" are the action functionals written for modulus of the complex time $\tau$.
For normalization requirement under such conditions we, obtain
\begin{multline*}
\idotsint\limits_{\Omega^{I}}\biggl|\Psi_{\tau}(x,X^{1}, \dots ,X^{n})\biggl|^{2}\,dx\,dX^1\cdots\,dX^N+\\
+\idotsint\limits_{\Omega^{II}}\biggl|\Psi_{\tau}(x,X^{1}, \dots ,X^{n})\biggl|^{2}\,dx\,dX^1\cdots\,dX^N= const.
\end{multline*}
Taking into account \eqref{eq:math:ex10} and \eqref{eq:math:ex11} we have
\begin{multline*}
\exp\biggl({\frac{2}{\hbar}U_{a}\varepsilon}\biggl)\idotsint\limits_{\Omega^{I}}\Biggl|\idotsint\limits_{\Omega^{I}}\Biggl(\int\exp{\biggl(-\frac{1}{\hbar}s_{34}[\gamma^{I}]\biggl)}\,[d\gamma^{I}]\Biggl)\delta (X_{4}^{k}- X_{3}^{k})\times\\
\times\Psi_{\tau_{3}}(x,X^{1}, \dots ,X^{n})\,dx_{3}\,dX_{3}^1\cdots\,dX_{3}^N\Biggl|^{2}\,dx_{4}\,dX_{4}^1\cdots\,dX_{4}^N+\\
+\idotsint\limits_{\Omega^{II}}\Biggl|\idotsint\limits_{\Omega^{II}}\Biggl\{\int\exp\biggl(-\frac{1}{\hbar}s_{34}[\gamma^{II}]\biggl)\times\\ \times\idotsint\Biggl(\prod\limits_{j=1,j\neq k}^{N}\exp{\biggl(-\frac{1}{\hbar}S_{3,4}^{j}[\Gamma^{j}]}\biggl) \,[d\Gamma^{j}]\Biggl)\,[d\gamma^{II}]\Biggl\}\times\\
\times\Psi_{\tau_{3}}(x,X^{1}, \dots ,X^{n})\,dx_{3}\,dX_{3}^1\cdots\,dX_{3}^N\Biggl|^{2}\,dx_{4}\,dX_{4}^1\cdots\,dX_{4}^N= const.
\end{multline*}
The integrals in both terms of the last expression are commensurable quantities having a microscopic order of magnitude. Therefore the first term much exceeds the second due to the macroscopic order of magnitude of the exponent entering in this term. Therefore it is the fist term determines the wave function normalization, i.e. the wave function of the particle-object localizes in the volume of the grain $k$ and at the instant of time $t_{4}$ it has the form \footnote{As we assigned above, the grain size be so small that its position in space can be determined by the unique coordinate $Y^{j}$.}
\begin{displaymath}
\Psi_{\tau_{4}}(x_{4})=\delta\left(x_{4}-Y(\tau_{4})\right),
\end{displaymath}
 Then for the real time we have
\begin{displaymath}
\Psi_{t_{4}}(x_{4})=\delta\left(x_{4}-Y(t_{4})\right).
\end{displaymath}
Thus, considering here model results in wave function collapse in the form of space localization. Actually, as soon as wave function of the particle object localizes, its interaction with the grains having numbers $j\neq k$ ceases \footnote{In accordance with the assumption of small instantaneous radius} and, in general, nonlocal particle-object visualizes macroscopically in the form of the single mass point.

\begin {center} \textbf{\textbf{5. CONCLUSION}}\\
\end {center}

Both types of the quantum object dynamics --- the Schro\"dinger evolution and the collapse --- is described by the integral wave equation with the kernel in the form of a path integral. The Schro\"dinger dynamics is determined by individual points motion along the virtual path (ultima analysi it is caused by gradient of the potential energy~\cite{bib:F}). This is a local deterministic process. The collapse dynamics is determined by the measure change of virtual path collection with the time in a local region of space (it is caused by time derivative of the potential energy). This change instantly effects on the wave function values in all space due to the wave function normalization conserving in the quantum processes \footnote{The conservation expresses entirety of the quantum object}. Thus the collapse is nonlocal physical process, what, however, does not conflict with relativity theory. Really, we have instantaneous causal connection between special remote events inside entire quantum object and there is no nonlocal interaction between different physical objects. If we take the integral wave equation with the kernel in the form of a path integral as a postulate, then this specific quantum process is the consequence of it, as well as the Schro\"dinger evolution.

Instant localization of the wave function in the measurement allows us to interpret the wave function as a mathematical direct image of the real continuum (in our case the quantum particle) \footnote{Instant spatial localization of quantum particles removes the famous contradiction between the spreading of the wave packet and registration of their as a point particles }. Thus, any quantum object is a continuum (set of the continuous media), which has specific properties, fully described by the integral wave equation with the kernel in the form of a path integral. This statement can be considered as the only postulate, determining any non-relativistic mechanical properties of quantum objects \footnote{spin is not considered here}. In such interpretation, the wave function can not be represented as a superposition of stationary states only. Quantum objects can be visualized at the macroscopic level, when they are in stationary states, but this does not exclude the possibility of their existence in the states differing from stationary, i.e. quantization phenomenon inherent only collection of the observable states.

Born's rule determines the rate of energy transfer from the particle-object to the grains. Grains are the macroscopic objects and, therefore, their physical characteristics have a random spread. The random spread of the physical quantities, that determine time of the macroscopic measuring process initiation in the grain, generates the probabilistic form of this rule.

The local realm principle does not violate in the reduction process. Actually, in  this case, we have not the correlation between spatially remoted events, we have the single event of spatial localization of the nonlocal object occurring simultaneously in the whole space. This, of course, violates Bell's inequality related to different events.

\vfill\eject

\end{document}